\documentclass[onecolumn]{article}
\usepackage{graphicx}
\usepackage{amsmath}
\usepackage{amsfonts}
\usepackage{amssymb}

\begin{document}

\title{Dynamics of the Time Horizon Minority Game}
\author{Michael L. Hart, Paul Jefferies and Neil F. Johnson\\Physics Department, Oxford University, Oxford OX1 3PU, U.K.}
\maketitle
\begin{abstract}
We present exact analytic results for a new version of the Minority Game (MG)
in which strategy performance is recorded over a finite time horizon. The
dynamics of this Time Horizon Minority Game (THMG) exhibit many distinct
features from the MG and depend strongly on whether the participants are fed
real, or random, history strings. The THMG equations are equivalent to a
Markov Chain, and yield exact analytic results for the volatility given a
specific realization for the quenched strategy disorder.
\end{abstract}

\bigskip\bigskip

\section{Introduction}

Agent-based models of complex adaptive systems are attracting significant
interest across many disciplines\cite{econophysics}. Typically each agent has
access to a limited set of recent global outcomes of the system; she then
combines this information with her limited strategy set chosen randomly at the
start of the game (i.e. quenched disorder) in order to decide an action at a
given timestep. These decisions by the $N$ agents feed back to generate the
fluctuations in the system's output. The Minority Game (MG) introduced by
Challet and Zhang\cite{challet,savit} offers possibly the simplest paradigm
for such a complex, adaptive system and has therefore been the subject of
intense research activity \cite{econophysics}-\cite{heimel}. Most studies of
the MG have focussed on a calculation of both time and configuration (i.e.
quenched disorder) averaged quantities, in particular the `volatility'
$\sigma$ where $\sigma$ is the standard deviation of the fluctuations. Our own
work has shown that the variations of this averaged $\sigma$ with memory size
$m$ can be quantitatively explained in terms of `crowd-anticrowd' collective
behavior \cite{us,crowd}. This crowd-anticrowd theory, which implicitly
assumes both time-averaging and configuration-averaging, is simple and
intuitive yet it yields useful analytic expressions \cite{us,crowd}. In terms
of more detailed microscopic theories, two complementary spin-glass approaches
have been shown to be remarkably powerful\cite{challet,heimel}.

In this paper, we wish to focus on the dynamics of the multi-agent game for a
given realization of the quenched disorder of initially picked strategies. We
present exact analytic results for a finite time horizon version of the
Minority Game, the Time Horizon Minority Game (THMG), in which strategy
performance is only recorded over the previous $\tau$ timesteps. This
seemingly trivial modification of the basic MG yields a dynamical system with
surprisingly rich dynamics. These dynamics depend strongly on whether the
participants are fed real (as opposed to random) history strings, and on the
nature of the quenched disorder corresponding to initial conditions. We
present a set of equations describing the THMG - these equations are
equivalent to a Markov Chain. This Markov Chain is used to generate accurate
analytic results for the resulting volatility in the THMG. Throughout the
paper, similarities and differences between the THMG and MG are pointed out
where appropriate. Section 2 provides a brief introduction to the MG and
provides exact analytic expressions for the volatility $\sigma$ for a given
configuration of quenched disorder. Section 3 discusses the THMG and provides
corresponding formulae for this game. Section 4 compares the analytic and
numerical results for the THMG. Section 5 provides the conclusion.

\section{The basic Minority Game (MG)}

The MG \cite{econophysics,challet} comprises an odd number of agents $N$ (e.g.
traders or drivers) choosing repeatedly between option 0 (e.g. buy or choose
route 0) and option 1 (e.g. sell or choose route 1). The winners are those in
the minority group; e.g. sellers win if there is an excess of buyers, drivers
choosing route 0 encounter less traffic if most other drivers choose route 1.
The outcome at each timestep represents the winning decision, 0 or 1. A common
bit-string of the $m$ most recent outcomes is made available to the agents at
each timestep \cite{memory}. The agents randomly pick $s$ strategies at the
beginning of the game, with repetitions allowed (quenched disorder). Each
strategy is a bit-string of length $2^{m}$ which predicts the next outcome for
each of the $2^{m}$ possible histories. After each turn, the agent assigns one
(virtual) point to each of her strategies which would have predicted the
correct outcome, and penalizes a strategy by one (virtual) point if it
incorrectly predicts the outcome. At each turn of the game, the agent uses the
most successful strategy, i.e. the one with the most virtual points, among her
$s$ strategies.

The number of agents holding a particular combination of strategies can be
written as a $D\times D\times\dots$ ($s$ terms) dimensional tensor $\Omega$,
where $D$ is the total number of available strategies. For $s=2$, this is
simply a $D\times D$ matrix where the entry $(i,j)$ represents the number of
agents who picked strategy $i$ and then $j$. The strategy labels are given by
the decimal representation of the strategy plus unity, for example the
strategy 0101 for $m=2$ has strategy label 5+1=6. This quenched disorder
$\Omega$ is fixed at the beginning of the game and can be written using the
full or reduced strategy spaces \cite{challet}. The value of $a_{R}^{\mu}$
\cite{challet} describes the prediction of strategy $R$ given the history
$\mu$, where $\mu$ is the decimal number corresponding to the $m$-bit binary
history string. Hence $a_{R}^{\mu}=-1$ denotes a prediction of choice `0'
while $a_{R}^{\mu}=1$ denotes a prediction of choice `1'). The approach of our
(time and configuration-averaged) crowd-anticrowd theory was to partition the
$N$ agents into $g$ groups in such a way that the groups themselves were
uncorrelated. This was achieved by considering the reduced strategy space,
which produces essentially identical results to the full strategy space for
the volatility. Specifically, each group $g$ contains a crowd of $n_{g}^{C}$
agents using highly correlated strategies, and an anticrowd $n_{g}^{A}$ of
agents using strategies which are highly anti-correlated to the crowd. This
leaves an effective super-agent of size $n_{g}=n_{g}^{C}-n_{g}^{A}$
representing each group. Since the strong correlations have now been removed
from the problem, the resulting groups are essentially uncorrelated and can be
considered as executing random walks with respect to each other in terms of
their decisions. As shown in Refs. \cite{us,crowd}, summing the resulting
variances yields excellent agreement with the numerical values for the
volatility $\langle\langle\sigma\rangle\rangle_{\Omega}$, where $\langle
\langle\dots\rangle\rangle_{\Omega}$ denotes averaging over all initial
configurations of quenched disorder $\{\Omega\}$.

Here we are interested instead in the detailed dynamics of the game for a
given choice of initial quenched disorder $\Omega$, hence we follow a more
microscopic approach. This is particularly relevant if the intended
application of such games is to understand financial markets, since such
markets should each correspond to just \emph{one} realization of the game
given an initial $\Omega$. Hence we imagine that a particular $\Omega$ has
already been chosen. Since the game involves a coin-toss to break ties in
strategy scores, this stochasticity also means that different runs for a given
$\Omega$ will also differ - we return to this point below. The number of
traders making decision $1$ (the `instantaneous crowd') minus the number of
traders making decision $0$ (the `instantaneous anticrowd') defines the net
`attendance' $A[t]$ at a given timestep $t$ of the game. This attendance
$A[t]$ is made up of two groups of traders at any one timestep: there are
$A_{D}[t]$ traders who act in a `decided' way, i.e. they do not require the
toss of a coin to decide which choice to make - this is because they have one
strategy that is better than their others, or because their highest-scoring
strategies are tied \emph{but} give the same response as each other to the
history $\mu_{t}$ at that turn of the game. In addition, there are $A_{U}[t]$
traders who act in an `undecided' way, i.e. they require the toss of a coin to
decide which choice to make - this is because they have two (or more)
highest-scoring tied strategies \emph{and} these give different responses to
the history $\mu_{t}$ at that turn of the game. Hence the outcome at timestep
$t$ is given by
\begin{equation}
A[t]=A_{D}[t]+A_{U}[t]\ \ \ .
\end{equation}
The state of the game at the beginning of timestep $t$ depends on the strategy
score vector ${\underline{s}}_{t}$\ and a history $\mu_{t}$ at that moment.
Henceforth we will drop the variable $t$ from the notation, but note that it
remains an implicit variable through the time-dependence of $\underline{s}$
and $\mu$. We also focus on $s=2$ strategies per agent, although the formalism
can be generalized in a straightforward way. At timestep $t$, $A_{D}$ is given
exactly by
\begin{equation}
A_{D}(\underline{s},\mu)=\sum_{R,R^{^{\prime}}=1}a_{R}^{\mu}(1+\mathrm{Sgn}%
[s_{R}-s_{R^{^{\prime}}}])\underline{\underline{\Psi}}_{R,R^{^{\prime}}}%
\end{equation}
where the symmetrized matrix $\underline{\underline{\Psi}}=\frac{1}%
{2}(\underline{\underline{\Omega}}+\underline{\underline{\Omega}}^{T})$ with
$\underline{\underline{\Omega}}$ representing the quenched disorder. The
element $\underline{\underline{\Omega}}_{R,R^{^{\prime}}}$ gives the number of
agents picking strategy $R$ and then $R^{\prime}$. The number of undecided
traders $N_{U}$ is given exactly by
\begin{equation}
N_{U}(\underline{s},\mu)=\sum_{R,R^{^{\prime}}=1}\delta(s_{R}-s_{R^{^{\prime}%
}})[1-\delta(a_{R}^{\mu}-a_{R^{^{\prime}}}^{\mu})]\underline{\underline
{\Omega}}_{R,R^{^{\prime}}}%
\end{equation}
and hence the attendance of undecided traders $A_{U}$\ is distributed
binomially in the following way:
\begin{equation}
A_{U}(\underline{s},\mu)\equiv\ 2\ \mathrm{Bin}[N_{U}(\underline{s},\mu
),\frac{1}{2}]-N_{U}(\underline{s},\mu)
\end{equation}
where the term `Bin' denotes a binomial distribution with $N_{U}(\underline
{s},\mu)$ trials and probability $1/2$.

The so-called `volatility' is used in finance to describe some statistical
characteristic of the fluctuations in the market. It does not have a unique
definition in the finance literature but is typically taken as some form of
`root-mean-square' fluctuation - however this definition leaves open the
question of \emph{which} mean should be computed. In the present context, it
makes sense to define the volatility in terms of the time-average of a
particular realization $k$ of the random process $A[t]$, given an initial
quenched disorder $\Omega$. As mentioned above, our crowd-anticrowd theory was
limited to consideration of a time-averaged volatility $\langle\langle
\sigma\rangle\rangle_{\Omega}$ which had \emph{also} been averaged over all
configurations of initial quenched disorder $\{\Omega\}$. The present work
goes beyond this limitation to consideration of a \emph{particular} choice of
quenched disorder $\Omega$. Consider any stochastic process $z(t)$ produced by
a particular realization $k$ of the game, given an initial quenched disorder
$\Omega$. This quantity $z(t)$ could represent the attendance $A[t]$ at
timestep $t$, or any time-dependent quantity derived from it such as the mean
attendance at timestep $t$ calculated over the past $n$ timesteps, or the
volatility defined as the root-mean-square attendance over the past $n$
timesteps. The finite time average of the $k$'th realization of this process
is given by
\begin{equation}
\lbrack^{(k)}z(t)]_{T}\equiv{\frac{1}{T}}\int_{t-T/2}^{t+T/2}\ ^{(k)}%
z(t^{\prime})dt^{\prime}\ \ .
\end{equation}
If $T$ is finite, then $[^{(k)}z(t)]_{T}$ is itself a random process. In real
financial markets, the volatility defined by such a finite-time average does
indeed fluctuate producing `volatility clustering'. Here we instead wish to
focus on the time-average which is defined in the $T\rightarrow\infty$ limit:
\begin{equation}
{\overline{^{(k)}z}}\equiv\mathrm{Lim_{\ T\rightarrow\infty}}\ \ {\frac{1}{T}%
}\int_{t-T/2}^{t+T/2}\ ^{(k)}z(t^{\prime})dt^{\prime}\ \ ,
\end{equation}
which no longer depends on $t$ or $T$ but in general does depend on the
particular realization $k$ of the ensemble that we have chosen in addition to
the dependence on the initial quenched disorder $\Omega$. In the absence of
stochastic tie-breaks, the attendance $A[t]$ would be deterministic hence
producing a deterministic Minority Game for a given initial quenched disorder
$\Omega$ \cite{paul}. However in the presence of stochastic tie-breaks, which
is the case of interest here, the game should self-average in the following
sense: for a given quenched disorder $\Omega$, the time-average of $A[t]$ or
any of its higher order correlation functions (e.g. volatility) for a given
realization $k$ should be equivalent to the ensemble average value taken over
all realizations $k$ at a given time $t$. We stress that this discussion is
for one particular (fixed ) quenched disorder $\Omega$ - we are \emph{not}
averaging over this quenched disorder. Henceforth we will therefore assume
that this ergodic principle holds given a particular $\Omega$, i.e. we assume
that the time-average and ensemble-average of both the attendance $A[t]$ and
the volatility are equivalent for fixed quenched disorder $\Omega$. We will
denote this average attendance as ${\overline{A}}$ and the associated average
volatility as $\sigma$, noting that both have an implicit dependence on the
initial quenched disorder $\Omega$. Hence for a given $\Omega$, the square of
the volatility is given exactly by the expectation value of the mean-square
attendance:
\begin{equation}
\sigma^{2}=\sum_{A}(A-{\overline{A}})^{2}P(A)
\end{equation}
where $P(A)$ is the probability that the attendance is given by $A$. From Eq.
(1) the attendance $A=A_{D}(\underline{s},\mu)+A_{U}(\underline{s},\mu,x)$
where we have included the stochastic variable $x$ to denote the coin-toss
process. Because this coin-toss process is unbiased, we have ${\overline
{A}_{U}(\underline{s},\mu,x)=0}$ and hence ${\overline{A}}={\overline{A}%
_{D}(\underline{s},\mu)}$. The probability $P(A)$ is exactly equivalent to the
probability of obtaining a given $\underline{s}$, $\mu$ and $x$. Since $x$ is
an independent variable, we have
\begin{equation}
P(A)=P(\mu|\underline{s})P(\underline{s})P(x)
\end{equation}
with $P(x)$ given by the binomial expression $^{N_{U}}C_{x}(\frac{1}%
{2})^{N_{U}}$. Here $P(\underline{s})$\ is the probability that the strategy
score vector is $\underline{s}$, while $P(\mu\mid\underline{s})$ is the
probability that the game is in a state where the history is $\mu$ given that
the strategy score vector is $\underline{s}$. Hence Eq. (7) can be rewritten
exactly as
\begin{equation}
\sigma^{2}=\sum_{\{\underline{s}\}}\bigg[\sum_{\{\mu\}}\bigg[\sum_{x=0}%
^{N_{U}}\bigg[\;^{N_{U}}C_{x}(\frac{1}{2})^{N_{U}}(A_{D}+2x-N_{U}%
-{\overline{A}})^{2}\bigg]P(\mu\mid\underline{s})\bigg]P(\underline{s})\bigg]%
\end{equation}
with the mean attendance being given exactly by
\begin{equation}
{\overline{A}}=\sum_{\{\underline{s}\}}\bigg[\sum_{\{\mu\}}\bigg[A_{D}%
\;P(\mu\mid\underline{s})\bigg]P(\underline{s})\bigg]%
\end{equation}
where in Eqs. (9) and (10) $A_{D}=A_{D}(\underline{s},\mu)$\ and $N_{U}%
=N_{U}(\underline{s},\mu)$. The difficulty in evaluating this expression for
the volatility in the MG, for a given quenched disorder $\Omega$, now lies in
the complexity of $P(\underline{s})$ and $P(\mu\mid\underline{s})$. As an
example of a specific realization of the initial quenched disorder, we shall
take $\Omega$ throughout this paper to be the following randomly-chosen matrix
in the reduced strategy space:
\begin{equation}
\underline{\underline{\Omega}}=\left(
\begin{array}
[c]{cccccccc}%
1 & 2 & 1 & 0 & 1 & 1 & 1 & 1\\
0 & 0 & 2 & 2 & 2 & 0 & 2 & 0\\
2 & 1 & 0 & 3 & 1 & 0 & 2 & 1\\
2 & 2 & 0 & 3 & 3 & 2 & 2 & 2\\
1 & 1 & 0 & 2 & 3 & 3 & 2 & 0\\
2 & 3 & 2 & 0 & 1 & 5 & 1 & 1\\
0 & 1 & 4 & 7 & 2 & 1 & 0 & 0\\
3 & 2 & 2 & 0 & 2 & 2 & 2 & 4
\end{array}
\right)  \ \ .
\end{equation}

Figure 1 shows the resulting $P(\underline{s})$ for $N=101$, $m=2$, $s=2$ and
$\Omega$ as shown above. The strategy scores themselves are written out
explicitly for the dominant score vectors. As can be seen, the probability
distribution $P(\underline{s})$ is uneven and has non-trivial structure.

\vskip\baselineskip

\section{Time Horizon Minority Game (THMG)}

In the basic MG, strategy scores are kept since the beginning of the game. One
might ask whether this is realistic in a `real-world' situation, given the
finite time-horizon under which most `agents' (e.g. traders) tend to operate.
Hence we will make a small modification to this rule - we introduce the Time
Horizon Minority Game (THMG) in which strategy scores are only kept over the
last $\tau$ turns of the game. Hence agents (e.g. traders) are limited to
assessing their strategies' performance over the last $\tau$ turns of the
game, in addition to the basic MG rule of viewing just the last $m$ steps of
the history. We focus on the low $m$ regime because of the richer dynamics,
however our formalism applies for all $m$.

Figure 2 shows numerical results for the variation in volatility as a function
of $\tau$, given different realizations of the initial quenched disorder
$\Omega$. Results are shown for $m=3$ using the full strategy space, and for
both real and random histories of the game. For real histories it can be seen
that the volatility is essentially periodic in $2.2^{m}$. This value
corresponds to the number of different paths in a De Bruijn graph linking all
$P=2^{m}$ histories. Such a graph is necessarily Eulerian since all the
vertex-degrees of a De Bruijn graph are even. Games that differ in $\tau$\ by
multiples of $2.2^{m}$\ show similar dynamics for $t>\tau$. The dynamics of
the THMG for $t<\tau$ are that of the basic MG, as can be seen in Figs. 3a and
3b for small $t$.

The peaks in Fig. 2 which arise at $\tau=2.2^{m}\lambda-1$ for real histories,
where $\lambda\geq1$ and is an integer, correspond to realizations of the game
that are purely deterministic. If the game performs a path around the de
Bruijn graph that is of a length $\tau$ and satisfies the condition that all
edges have been visited equally during the course of this path, then
$\underline{s}$ and $\mu$\ at the start of the cycle are identical to
$\underline{s}$ and $\mu$\ at the end of the cycle, i.e. $\Delta s_{cycle}=0$
and $\Delta\mu_{cycle}=0$. Note that even for large $\lambda$\ this is very
likely to happen due to the nature of the minority game in the efficient
regime \cite{paul}. In the THMG it is observed to occur very soon after
$t$\ becomes larger than $\tau$. Once this type of path occurs for integer
$\lambda$, the game evolves such that ${\underline{s}}^{\mu}$ is equal to
$\pm\underline{a}^{\mu}$ at any subsequent time-step of the game. Note that
for integer $\lambda$\ the strategies are scored for one time-step less than
the period of this special cycle; this fact together with $\Delta s_{cycle}=0$
and $\Delta\mu_{cycle}=0$ imply that ${\underline{s}}^{\mu}=\pm\underline
{a}^{\mu}.$ When ${\underline{s}}^{\mu}=\pm\underline{a}^{\mu}$, we are left
with the unique situation where all tied strategies that have a particular
score at history $\mu$ give the same game decision, i.e. $+1$ or $-1$. When
this happens there are no longer any traders that have tied strategies telling
them to make opposing decisions. None of the game dynamics for this turn of
the game are therefore decided by the tossing of coins - this turn is hence
purely deterministic. Once the game has found such a deterministic state, it
never leaves it and the game henceforth evolves such that $\underline{s}^{\mu
}=\pm\underline{a}^{\mu}$. Figure 3a shows the finite time-average standard
deviation (over $100$ turns) of the attendance, together with the actual
number of traders making a given decision as a function of time. The case
shown corresponds to $\lambda=100$ and hence $\tau=1599$. For time-steps $0$
to $1599$ the system is equivalent to that of the basic MG, whereas from
$t=1600$ onwards the effect of the time horizon on strategy scores becomes
apparent. The system only takes about $40$ time-steps to become locked into
the deterministic state described above. We have observed at low $m$\ and for
several randomly selected $\Omega$, that once $t$ becomes larger than $\tau$
then the game rapidly finds the deterministic state described above. The game
dynamics hence become deterministic and periodic with period $\tau+1$. It is
possible to construct specially chosen $\Omega$\ matrices such that the above
special cycle is not found during a run of the corresponding game, however we
are here interested in characterising the dynamics for a `typical' (i.e.
randomly chosen) $\Omega$. Figure 3b shows the corresponding results for
$\tau=1600.$ We stress that the THMG with random histories has very different
dynamics as can be seen in Fig. 2.

We now present an analytic description of the THMG. Consider $\tau+1$
consecutive histories in the game and call this a $\mu$-path, denoted as
$\mu_{path}.$ We define $\underline{\underline{G}}$\ as the matrix telling us
which transitions between histories are allowed in the game - the matrix
$\underline{\underline{G}}$ is the $P\times P$ `adjacency' matrix of a de
Bruijn graph of order $m$. The element $\underline{\underline{G}}_{\mu,\nu}$
has value $1$ if history $\nu$ can follow history $\mu$ in the game, but has
value $0$\ if the transition is disallowed. The adjacency matrix is hence
given by
\begin{equation}
\underline{\underline{G}}_{\mu,\nu}=\delta(2\mu\%P-\nu)+\delta(2\mu\%P+1-\nu)
\end{equation}
where $A\%B$\ is the remainder when $A$ is divided by $B$. Here $\mu$ and
$\nu$ denote the nodes (i.e. histories) in the de Bruijn graph where $\mu
,\nu=0,1,..$. The matrix $\underline{\underline{G}}$ quantifies which paths
around history space are allowed and hence we can write down a rule for
determining which $\mu_{path}$ transitions are permitted in the game. Let
$\mu_{path}(t-1)=\mu_{t-\tau-1}\rightarrow\mu_{t-\tau}\rightarrow...\mu_{t-1}%
$. Whether $\mu_{path}(t-1)$ actually arises in the game, and whether the
transition to $\mu_{path}(t)=\mu_{t-\tau}\rightarrow\mu_{t-\tau+1}%
\rightarrow...\mu_{t}$ is allowed, depends on whether all the transitions are
allowed between histories in $\mu_{path}(t-1)$\ and the corresponding history
in $\mu_{path}(t)$. Figure 4 shows an example of the allowed transitions
between $\mu$-paths for $m=2,\tau=2.$ Let us define a scalar quantity $\Gamma$
given by
\begin{equation}
\Gamma=\prod_{i=0}^{\tau}\underline{\underline{G}}_{\mu_{t-1-i},\mu_{t-i}%
}\ \ .
\end{equation}
The transition between $\mu_{path}(t-1)$ and $\mu_{path}(t)$ is allowed if
$\Gamma=1$, but is disallowed if $\Gamma=0$. The increment in score vector
when passing from one history at time-step $t-1$ to the next history at
time-step $t$ is given by
\begin{equation}
\underline{\delta s}_{\mu_{t-1}\rightarrow\mu_{t}}=(2\underline{a}^{\mu_{t-1}%
}-1)(2(\mu_{t}\%2)-1)
\end{equation}
It follows that
\begin{equation}
\underline{s}_{\mu_{path}}=\sum_{i=0}^{\tau-1}\underline{\delta s}%
_{\mu_{t-1-i}\rightarrow\mu_{t-i}}\ \ .
\end{equation}

We can evaluate the exact number of decided and undecided traders for a given
$\mu_{path}$, and we can also identity which $\mu_{path}$ are allowed in the
game; hence we can find the transition matrix giving the probability that a
particular $\mu_{path}$ at an arbitrary timestep evolves to the next allowed
$\mu_{path}$ with final history $\mu_{t}$. This transition matrix is given as
follows:
\begin{align}
\underline{\underline{T}}_{\mu_{path}(t-1),\mu_{path}(t)}  & =\sum
_{x=0}^{N_{U}}\bigg[\;^{N_{U}}C_{x}(\frac{1}{2})^{N_{U}}\delta\bigg
[\mathrm{Sgn}(A_{D}+2x-N_{U})\nonumber\\
& +(2(\mu_{t}\%2)-1)\bigg]\bigg]%
\end{align}
where $A_{D}=A_{D}(\mu_{path}(t-1))=A_{D}(\underline{s}_{\mu_{path}(t-1)}%
,\mu_{t-1})$\ and $N_{U}=N_{U}(\mu_{path}(t-1))=N_{U}(\underline{s}%
_{\mu_{path}(t-1)},\mu_{t-1})$. The size of $\underline{\underline{T}}$\ is
given by $\phi=2^{(m+\tau)}$. For the basic MG $\phi$ would grow indefinitely
with time $t$, however for the THMG it is of fixed size. Having obtained this
transition matrix for the THMG then allows us to calculate various macroscopic
quantities, in particular $P(\mu_{path})$. The vector $\underline{P(\mu
_{path})}$\ satisfies $\underline{P(\mu_{path})}=\underline{P(\mu_{path}%
)}\underline{\underline{T}}$ which is the transpose of an
eigenvector-eigenvalue equation. The vector $\underline{P(\mu_{path})}$\ is
also a stationary distribution of the Markov Chain:
\begin{equation}
P(\mu_{path})=[\phi^{-1}\underline{1}\;\underline{\underline{T}}^{\infty
}]_{\mu_{path}}\ \ .
\end{equation}
The expression given for $P(\mu_{path})$\ represents an average over all games
with different $\mu$-paths defining the initial conditions of the game. This
is achieved by taking an average over all $\phi$\ rows of the matrix
$\underline{\underline{T}}^{\infty}$. This method assumes that, for all
possible states in the set $\{\mu_{path}\}$, if a state is visited then the
state can be re-visited during a run of the game. We note that if there are
closed irreducible subsets in the set of $\mu$-paths $\{\mu_{path}\}$\ which
are visited during the time evolution of the THMG, the game can then lock into
a deterministic state. This situation of deterministic states arises for
$\tau=2.2^{m}\lambda-1$ as discussed earlier. In this case, the present method
for calculating $P(\mu_{path})$ could be improved to account more fully for
the deterministic dynamics since in general the system is not ergodic We do
not consider such a calculation here, but note that the values of $\sigma$
obtained using the present method may still be very accurate (see later Fig. 6
for $\tau=7$ for example).

Figure 5a shows the measured $P(\mu_{path})$\ for $N=101$, $m=2$, $s=2$,
$\Omega$ as in Eq. (11) and $\tau=2$. Figure 5b shows the corresponding
calculated $P(\mu_{path})$ using Eq. (13). The agreement is excellent,
demonstrating that our expression for $P(\mu_{path})$ is exact.

\vskip\baselineskip

\section{Results}

We now show how to calculate the volatility of the THMG exactly using Eq. (9).
Without loss of generality, we replace both $P(\underline{s})$ and $P(\mu
\mid\underline{s})$ by the probability $P(\mu_{path})$ that the game has just
passed through one particular $\mu_{path}$ at a timestep $t$. This reduces the
problem of calculating the volatility $\sigma$ in the THMG to that of studying
a Markov Chain whose states are given by the set $\{\mu_{path}\}$. Note that
whilst the THMG is homogeneous, the basic MG is not since the size of the set
$\{\mu_{path}\}$ grows as $t$ increases. Replacing both $P(\underline{s})$ and
$P(\mu\mid\underline{s})$ by the probability $P(\mu_{path})$ in Eq. (9), we
have an equivalent exact expression for $\sigma$ given by :
\begin{equation}
\sigma^{2}=\sum_{\{\mu_{path}\}}\bigg[\sum_{x=0}^{N_{U}}\bigg[\;^{N_{U}}%
C_{x}(\frac{1}{2})^{N_{U}}(A_{D}+2x-N_{U}-{\overline{A}})^{2}\bigg]%
P(\mu_{path})\bigg]\ \ .
\end{equation}
In order to calculate $A_{D}$\ and $N_{U}$ exactly for a given $\mu_{path}$,
the score vector to be used is that which would be obtained if the game went
through a path of histories corresponding to $\mu_{path}$ as in Eq. (15); the
history to be used is the last history in the path $\mu_{path}$. Denoting this
last history as $\mu$, we have $A_{D}=A_{D}(\mu_{path})=A_{D}(\underline
{s}_{\mu_{path}},\mu)$ and $N_{U}=N_{U}(\mu_{path})=N_{U}(\underline{s}%
_{\mu_{path}},\mu)$. We note that the exact analytic expression given for
$\sigma$ in Eq. (18) applies to the limiting case where the number of
realizations used to numerically\ determine $\sigma$ tends to infinity (i.e.
ensemble average). We also note that using the formalism described, we could
alternatively have obtained a finite time-average volatility $\sigma$ between
$t=t_{1}$ and $t=t_{2}$ given a specific $\mu_{path}$ at $t=t_{1}$.

Using Eq. (17) in Eq. (18), together with Eqs. (2), (3) and (10), we obtain
analytic values for $\sigma$ in the THMG given the initial quenched disorder
$\Omega$. Figure 6 shows a comparison between the analytic values for the
volatility $\sigma$ and the numerical values taken from the game as a function
of the time-horizon $\tau$. Here $N=101$, $m=2$, $s=2$, and $\Omega$ is given
in Eq. (11). We only show $\tau=1\rightarrow2.2^{m}+1$, i.e. $\lambda=1$;
however similar results can also be obtained for $\lambda>1$. The agreement is
excellent, with the numerical and analytic lines essentially coincident. This
demonstrates the power and accuracy of the present Markov Chain formalism as
developed for the THMG.

\vskip\baselineskip

\section{Conclusion}

In summary, we have introduced and studied a finite time horizon version of
the Minority Game (THMG). We have presented exact analytic expressions for the
volatility in both the THMG and the basic MG, for a given configuration of
initial quenched disorder $\Omega$. We have presented an analytic theory to
describe the dynamics of the THMG by obtaining an analytic expression for the
transition matrix in terms of the set $\{\mu_{path}\}$. As an example of what
can be achieved analytically, we obtained excellent agreement between analytic
and numerical values for the THMG volatility, given knowledge of the initial
quenched disorder $\Omega$. Finally we would like to stress that our
theoretical approach and results avoid having to keep track of the labels of
individual agents - our results are obtained for a specific initial quenched
disorder $\Omega$, however this $\Omega$ describes many possible arrangements
of individual agents. In this sense, $\Omega$ defines a macrostate for which
there are many possible microstates corresponding to different initial
strategy choices by the $N$ individual agents.

\bigskip

We thank Pak Ming Hui for discussions.

\pagebreak

\pagebreak 

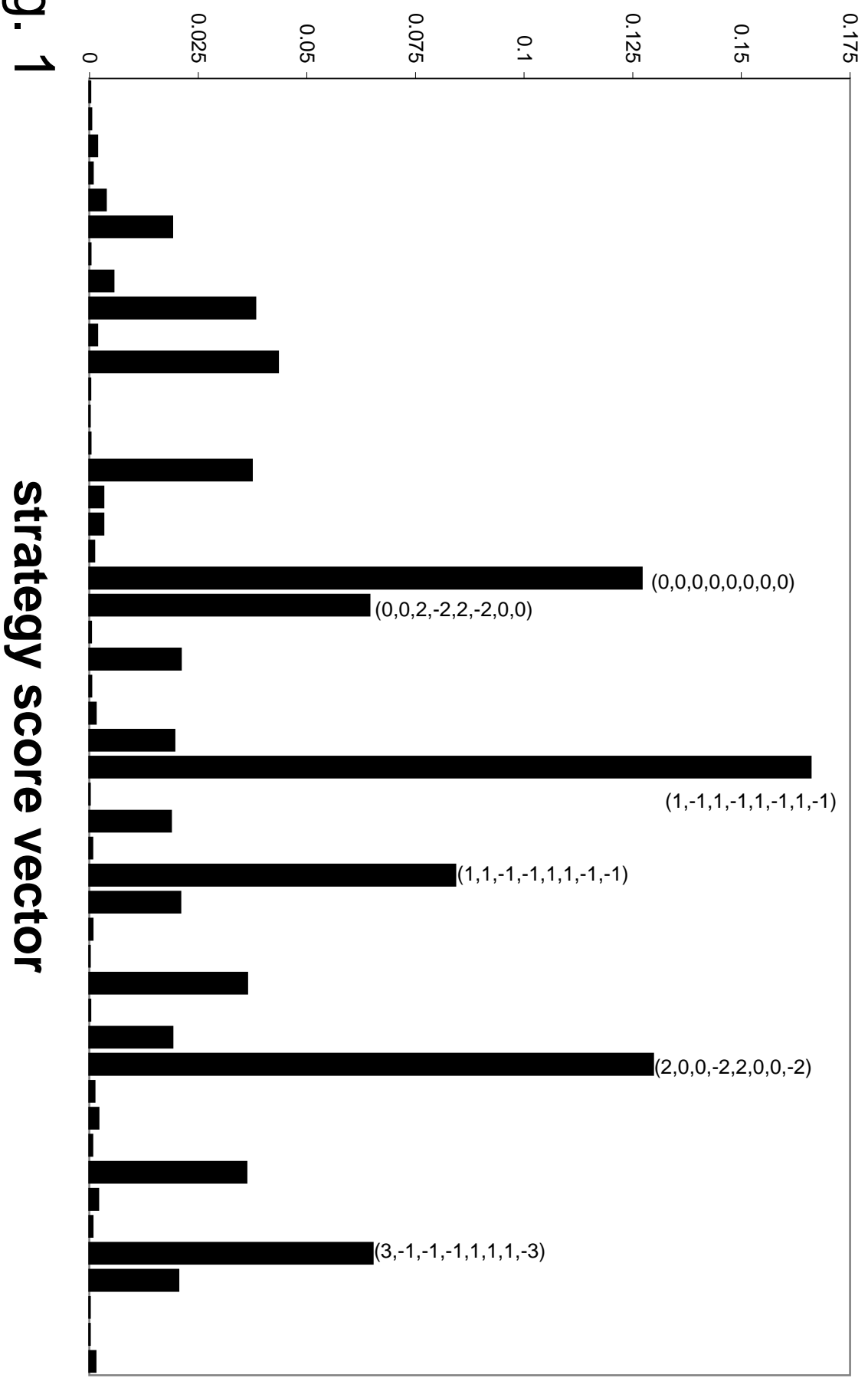
FIG. 1. The probability of occurrence $P(\underline{s})$ of the strategy score
vector obtained numerically from the Minority Game (MG). Strategy score
vectors \underline{$s$} are listed in an arbritary order along the x-axis.
Unless otherwise stated, the game parameters in Figs. 1-6 are as follows:
$N=101$, $m=2$, $s=2$ with the initial quenched disorder matrix $\Omega$ taken
from Eq. (11).

\bigskip

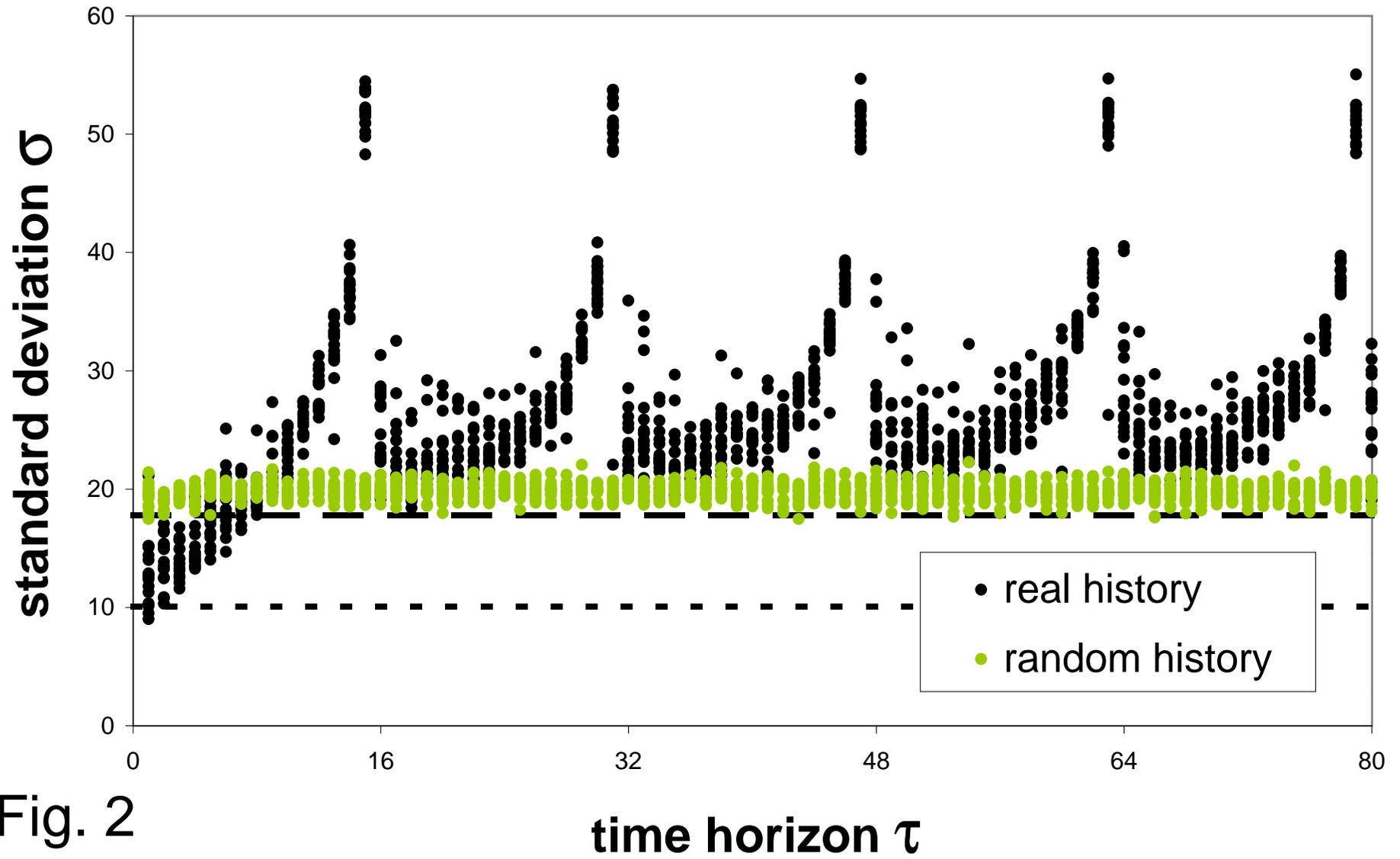
FIG. 2. Standard deviation (volatility) $\sigma$ as a function of time horizon
$\tau$ for the Time Horizon Minority Game (THMG) with $m=3$. Results are
obtained using the full strategy space. Numerical data is collected using real
histories (black circles) and random histories (grey circles) with randomly
selected quenched disorder matrices $\Omega$. The lower short-dashed line
shows the value of the volatility for the basic MG in the high $m$ limit. The
upper long-dashed line shows the configuration-average volatility (i.e.
average over quenched disorder $\Omega$) for the basic MG with $m=3$.

\bigskip

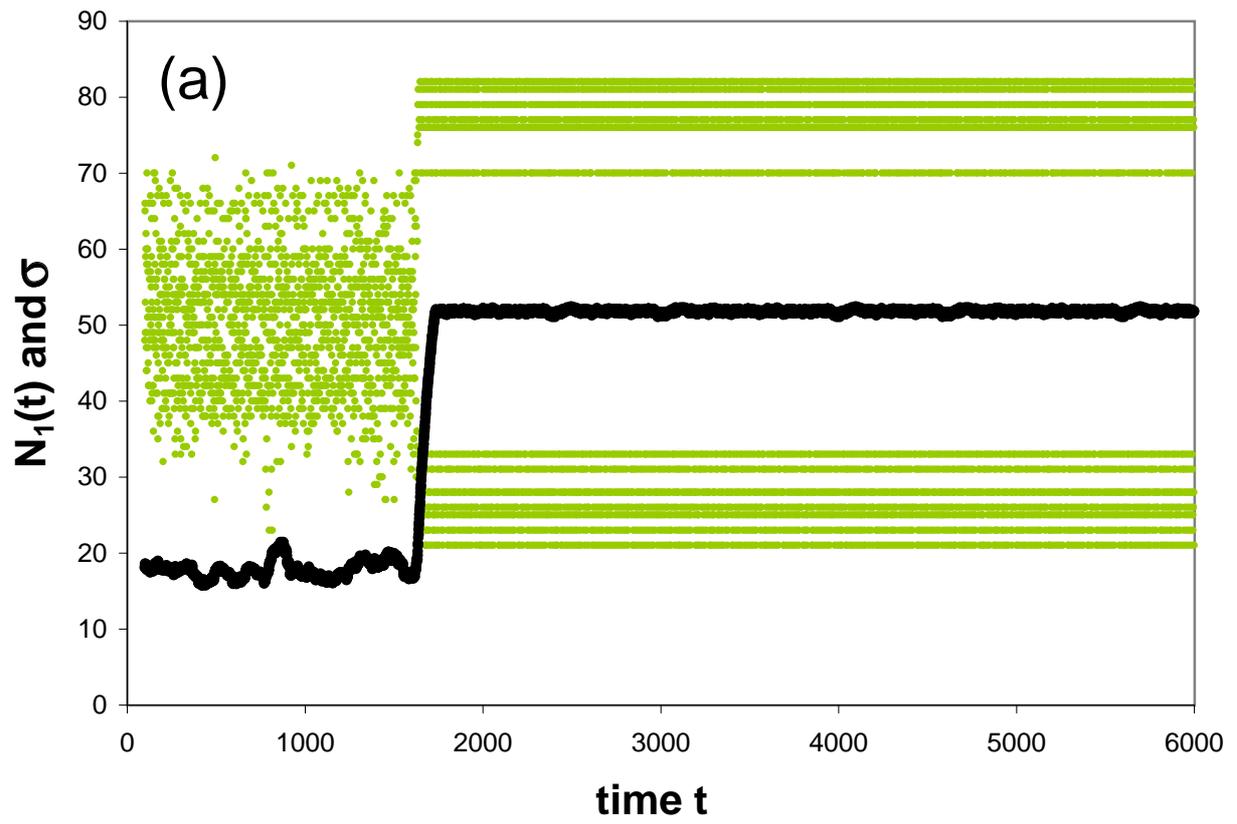
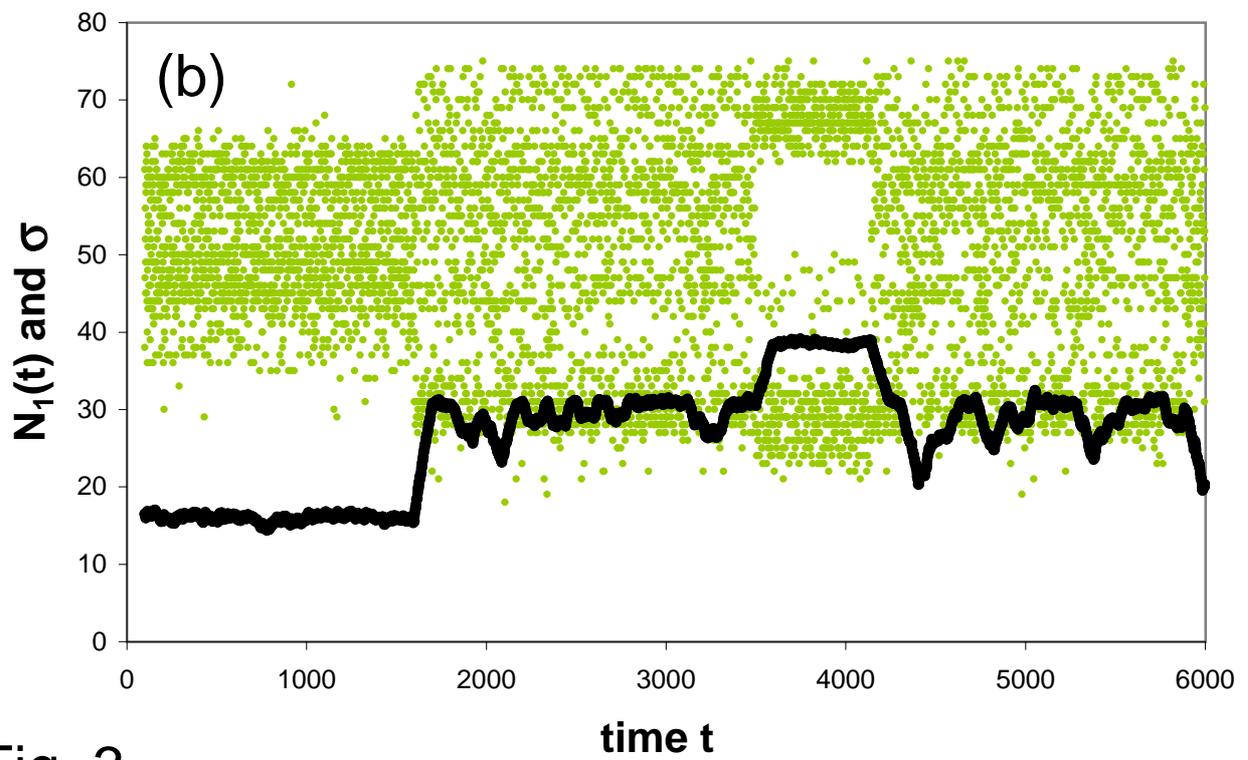
FIG. 3. The finite time-average standard deviation (average taken over $100$
turns) of the attendance of traders (black line) together with the number of
traders choosing `1' (grey circles) as a function of time $t$. (a)
$\lambda=100$, i.e. $\tau=1599$. (b) $\tau=1600$.

\bigskip

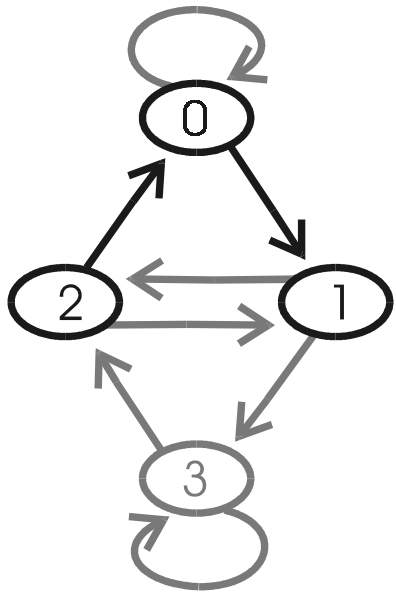
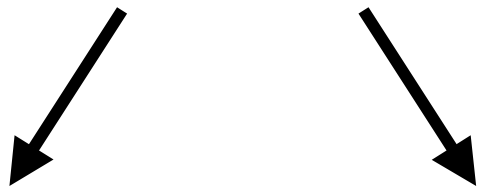
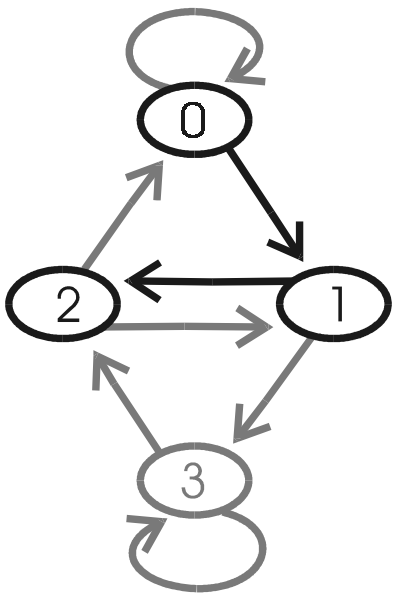
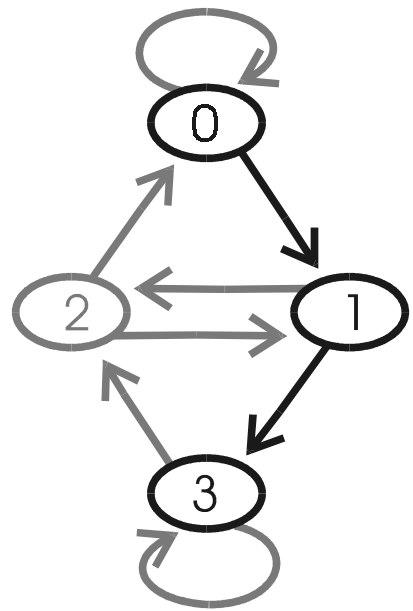
FIG. 4. Schematic diagram showing the allowed transitions between $\mu$-paths
in the THMG for the following example: $\mu_{path}(t-1)=2\rightarrow
0\rightarrow1$ and $m=2,\tau=2.$

\bigskip

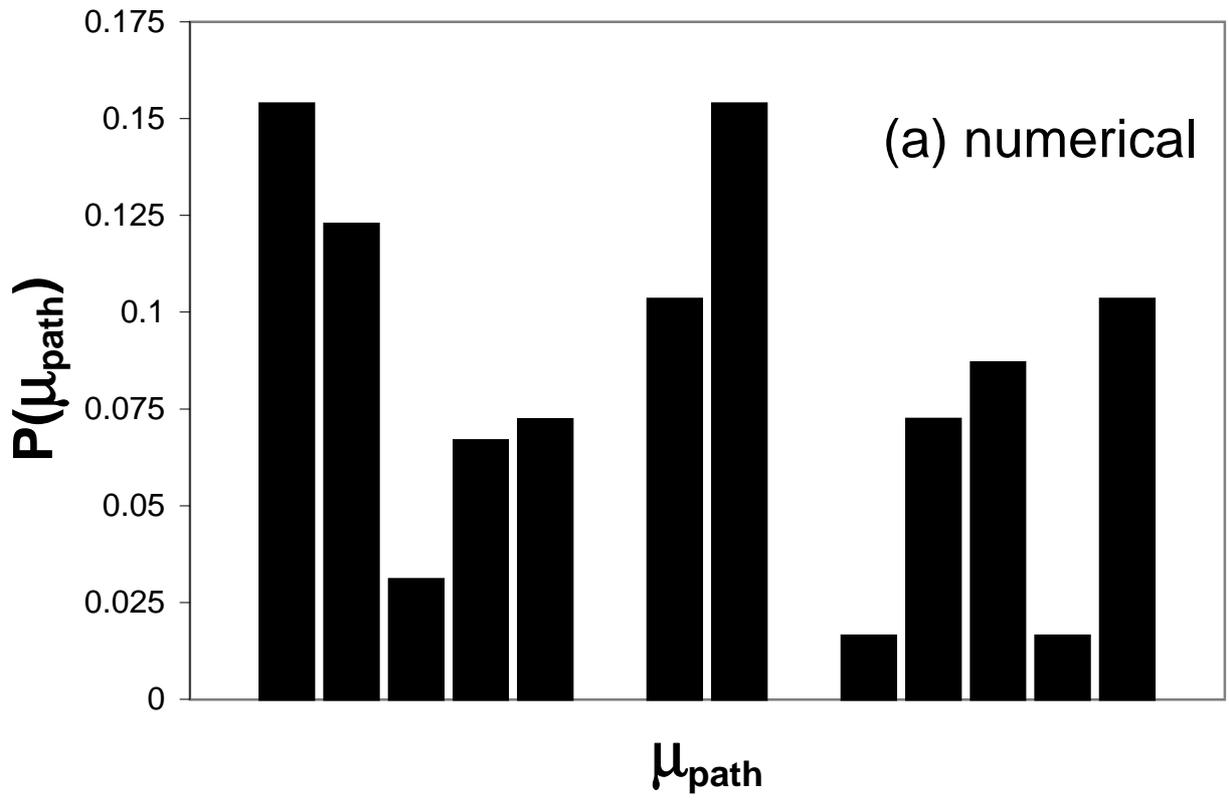
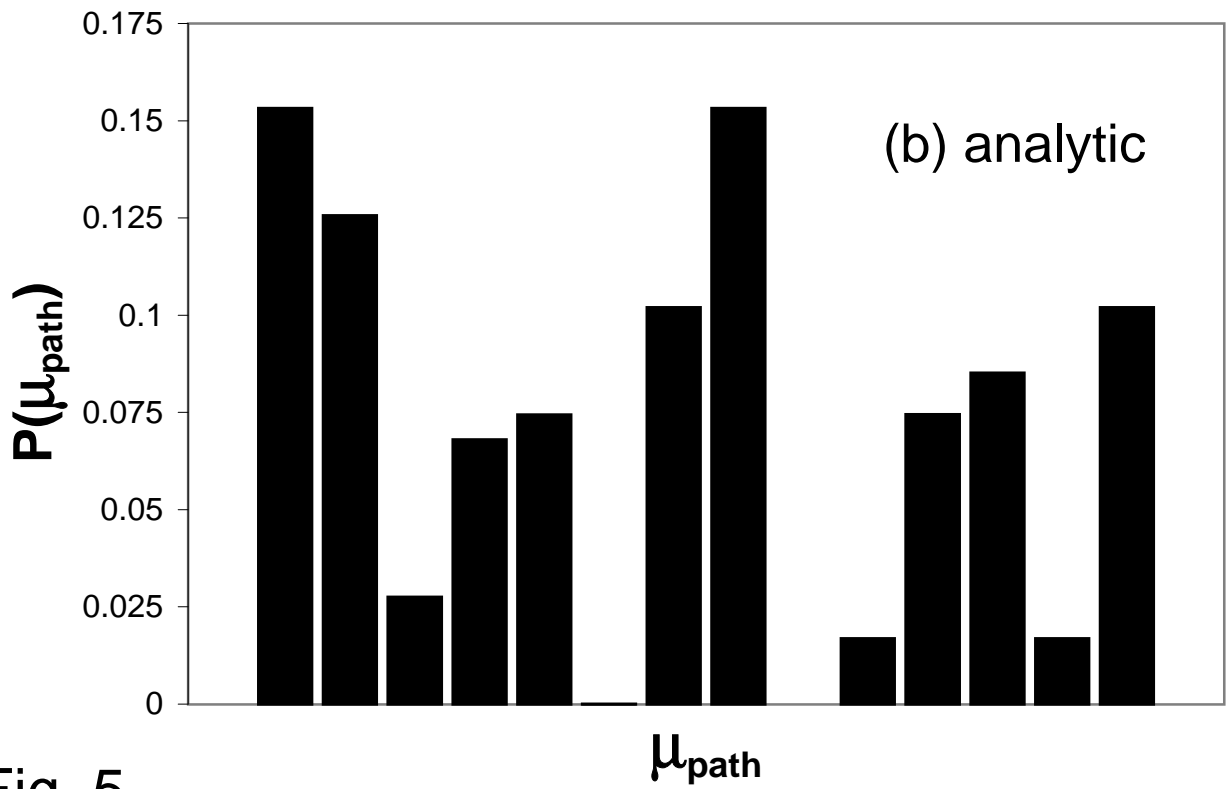
FIG. 5. (a) Numerical and (b) analytic results for the probability
distribution $P(\mu_{path})$ as a function of $\mu_{path}$. $\mu$-path are
listed in order of increasing decimal representation along the x-axis. Game
parameters are $N=101$, $m=2$, $s=2$, $\Omega$ as in Eq. (11), and $\tau=2$.

\bigskip

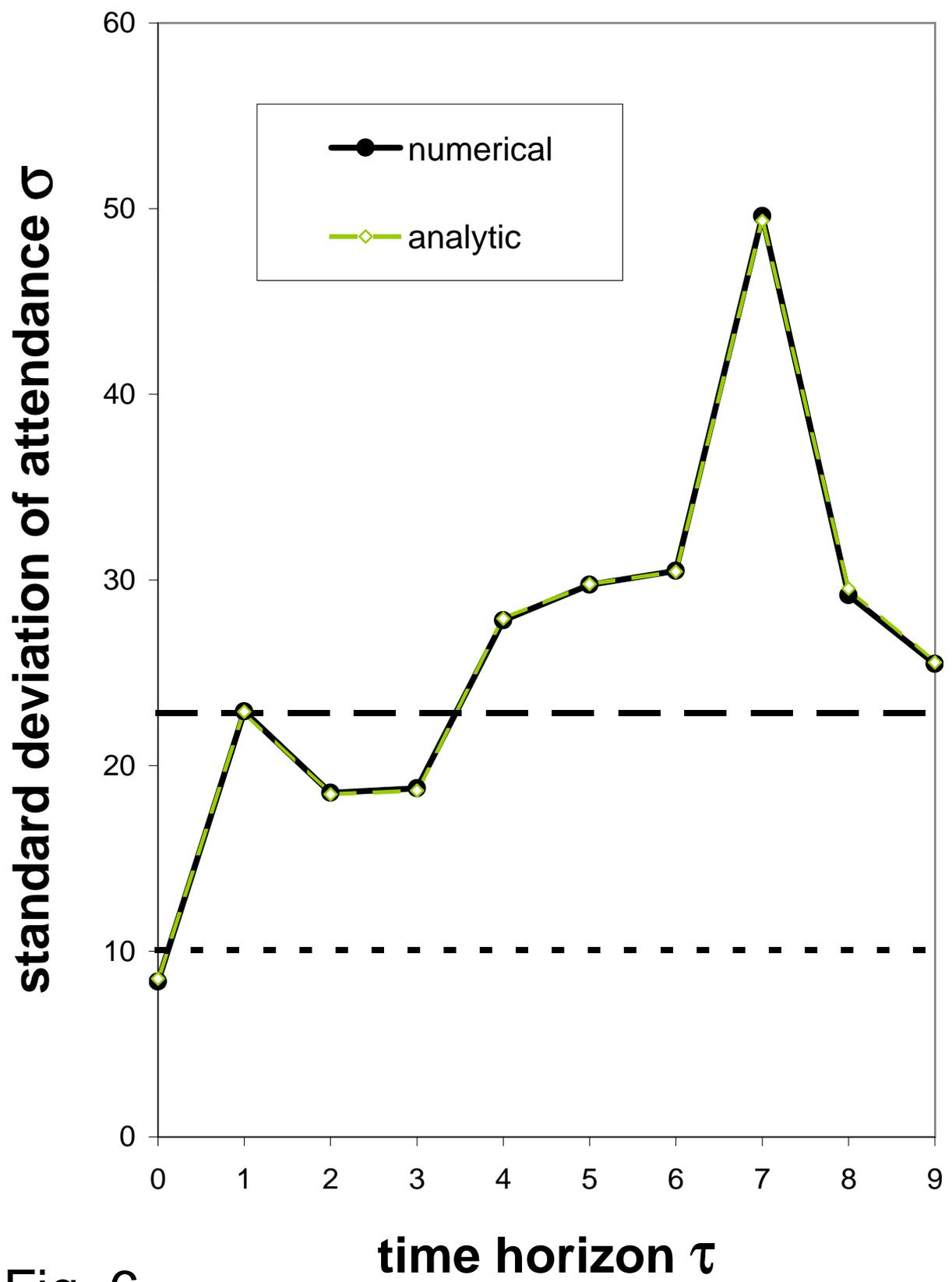
FIG. 6. Standard deviation (volatility) $\sigma$ for the THMG as a function of
time horizon $\tau$, for a given initial quenched disorder $\Omega$. Exact
analytic values are grey diamonds joined by a thick grey dashed line,
numerical values taken from the game simulation are black circles joined by a
thick black line. These lines are essentially coincident thereby demonstrating
the excellent agreement between the analytic theory and the numerical results.
Here $N=101$, $m=2$, $s=2$, and $\Omega$ is from Eq. (11). Lower short-dashed
line shows the value of the volatility for the basic MG in the high $m$ limit.
Upper long-dashed line shows the configuration-average volatility (i.e.
average over quenched disorder $\Omega$) for the basic MG with $m=2$.

\newpage

\begin{thebibliography}{99}
\bibitem{econophysics}See http://www.unifr.ch/econophysics for a detailed
account of previous work on agent-based games such as the Minority Game.

\bibitem{challet}D. Challet and Y.C. Zhang, Physica A \textbf{246}, 407
(1997); \emph{ibid.} \textbf{256}, 514 (1998); \emph{ibid.} \textbf{269}, 30
(1999); D. Challet and M. Marsili, Phys. Rev. E \textbf{60}, R6271 (1999); D.
Challet, M. Marsili, and R. Zecchina, Phys. Rev. Lett. \textbf{84}, 1824
(2000); M Marsili, D. Challet and R. Zecchina cond-mat/9908480; M Marsili and
D. Challet cond-mat/0102257.

\bibitem{savit}R. Savit, R. Manuca and R. Riolo, Phys. Rev. Lett. \textbf{82},
2203 (1999). See also Physica A \textbf{276}, 234 (2000) and 265 (2000).

\bibitem{dHR}R. D'Hulst and G.J. Rodgers, Physica A \textbf{270}, 514 (1999).

\bibitem{us}M. Hart, P. Jefferies, N.F. Johnson and P.M. Hui,
cond-mat/0003486; Phys. Rev. E \textbf{63}, 017102 (2001); cond-mat/0005152;
cond-mat/0008385 (to appear in Eur. J. Phys. B 2001); P. Jefferies, M. Hart,
N.F. Johnson and P.M. Hui, J. Phys. A: Math. Gen. \textbf{33} L409 (2000).

\bibitem{crowd}N.F. Johnson, P.M. Hui, D. Zheng and M. Hart, J. Phys. A: Math.
Gen. \textbf{32} L427 (1999); N.F. Johnson, M. Hart and P.M. Hui, Physica A
\textbf{269}, 1 (1999).

\bibitem{sherrington}A. Cavagna, J.P. Garrahan, I. Giardina and D.
Sherrington, Phys. Rev. Lett. \textbf{83}, 4429 (1999); J.P. Garrahan, E. Moro
and D. Sherrington, cond-mat/0012269; A. Cavagna, Phys. Rev. E \textbf{59},
R3783 (1999).

\bibitem{memory}See D. Challet and M. Marsili, cond-mat/0004196 for
discussions of the relevance of the actual memory in the MG, and diffusion
around de Bruijn graphs.

\bibitem{heimel}J.A.F. Heimel and A.C.C. Coolen, cond-mat/0012045.

\bibitem{paul}P. Jefferies, M.L. Hart and N.F. Johnson (in preparation).
\end{thebibliography}
\end{document}